\documentclass[12pt,epsf]{article}
\usepackage{graphicx}
\usepackage{epsfig}
\begin{document}

\def\a{\alpha}
\def\b{\beta}
\def\c{\varepsilon}
\def\d{\delta}
\def\e{\epsilon}
\def\f{\phi}
\def\g{\gamma}
\def\h{\theta}
\def\k{\kappa}
\def\l{\lambda}
\def\m{\mu}
\def\n{\nu}
\def\p{\psi}
\def\q{\partial}
\def\r{\rho}
\def\s{\sigma}
\def\t{\tau}
\def\u{\upsilon}
\def\v{\varphi}
\def\w{\omega}
\def\x{\xi}
\def\y{\eta}
\def\z{\zeta}
\def\D{\Delta}
\def\G{\Gamma}
\def\H{\Theta}
\def\L{\Lambda}
\def\F{\Phi}
\def\P{\Psi}
\def\S{\Sigma}

\def\o{\over}
\def\beq{\begin{eqnarray}}
\def\eeq{\end{eqnarray}}
\newcommand{\gsim}{ \mathop{}_{\textstyle \sim}^{\textstyle >} }
\newcommand{\lsim}{ \mathop{}_{\textstyle \sim}^{\textstyle <} }
\newcommand{\vev}[1]{ \left\langle {#1} \right\rangle }
\newcommand{\bra}[1]{ \langle {#1} | }
\newcommand{\ket}[1]{ | {#1} \rangle }
\newcommand{\EV}{ {\rm eV} }
\newcommand{\KEV}{ {\rm keV} }
\newcommand{\MEV}{ {\rm MeV} }
\newcommand{\GEV}{ {\rm GeV} }
\newcommand{\TEV}{ {\rm TeV} }
\def\diag{\mathop{\rm diag}\nolimits}
\def\Spin{\mathop{\rm Spin}}
\def\SO{\mathop{\rm SO}}
\def\O{\mathop{\rm O}}
\def\SU{\mathop{\rm SU}}
\def\U{\mathop{\rm U}}
\def\Sp{\mathop{\rm Sp}}
\def\SL{\mathop{\rm SL}}
\def\tr{\mathop{\rm tr}}

\def\IJMP{Int.~J.~Mod.~Phys. }
\def\MPL{Mod.~Phys.~Lett. }
\def\NP{Nucl.~Phys. }
\def\PL{Phys.~Lett. }
\def\PR{Phys.~Rev. }
\def\PRL{Phys.~Rev.~Lett. }
\def\PTP{Prog.~Theor.~Phys. }
\def\ZP{Z.~Phys. }

\newcommand{\beqr}{\begin{array}}  
\newcommand {\eeqr}{\end{array}}
\newcommand{\la}{\left\langle}  
\newcommand{\ra}{\right\rangle}
\newcommand{\non}{\nonumber}  
\newcommand{\ds}{\displaystyle}
\newcommand{\red}{\textcolor{red}}
\def\ubl{U(1)$_{\rm B-L}$}
\def\REF#1{(\ref{#1})}
\def\lrf#1#2{ \left(\frac{#1}{#2}\right)}
\def\lrfp#1#2#3{ \left(\frac{#1}{#2} \right)^{#3}}
\def\OG#1{ {\cal O}(#1){\rm\,GeV}}


\baselineskip 0.7cm

\begin{titlepage}

\begin{flushright}
UT-09-02\\
IPMU 10-0029
\end{flushright}

\vskip 1.35cm
\begin{center}
{\large \bf
Semi-direct Gauge Mediation in Conformal Windows of Vector-like Gauge Theories
}
\vskip 1.2cm
T. T. Yanagida$^{1,2}$ and Kazuya Yonekura$^{1,2}$ 
\vskip 0.4cm

{\it $^1$ Institute for the Physics and Mathematics of 
the Universe (IPMU),\\ 
University of Tokyo, Chiba 277-8568, Japan\\
$^2$  Department of Physics, University of Tokyo,\\
    Tokyo 113-0033, Japan}

\vskip 1.5cm

\abstract{ 
Direct gauge mediation models using the Intriligator-Seiberg-Shih (ISS) metastable vacua suffer from the Landau pole problem of the standard model
gauge couplings and the existence of R symmetry forbidding gaugino masses. These problems may be solved by using the recently proposed
SUSY breaking models in a conformal window of the vector-like $SU(N_C)$ gauge theory with gauge singlets.
In this paper we propose a model of gauge mediation based on the SUSY-breaking model in the conformal window, and study the dynamics 
for the SUSY breaking. 
In the model, there are massive vector-like bifundamental fields charged under both $SU(N_C)$
and the standard model gauge group, and our model can be regarded as a semi-direct gauge mediation model. The color number $N_C$ can be small to avoid
the Landau pole problem, and the R symmetry is also broken under a reasonable assumption on the strong dynamics of the model.
The model possesses only one free parameter, and the gaugino and sfermion masses are naturally of the same order.
}
\end{center}
\end{titlepage}

\setcounter{page}{2}

\section{Introduction}
\label{sec:1}
It is well known \cite{Intriligator:2006dd} that supersymmetric (SUSY) $SU(N_C)$ gauge theories with $N_F$ pairs of massive quarks and antiquarks, 
$Q$ and ${\tilde Q}$, have SUSY-breaking metastable vacua for $N_C+1\le N_F <\frac{3}{2}N_C$.
In the SUSY-breaking vacua, the direct gauge mediation  
takes place naturally if some pairs of quarks and antiquarks
carry the standard-model (SM) gauge charges by embedding the $SU(5)_{\rm GUT}$ gauge group into a subgroup of the flavor $SU(N_F)$
(see Ref.~\cite{Kitano:2010fa} and references therein).
From the point of view of the electric description of the $SU(N_C)$ theory, the fields charged under $SU(5)_{\rm GUT}$ have a mass term,
$W=mQ{\tilde Q}$, in the superpotential, and the model resembles a semi-direct gauge mediation~\cite{Izawa:1997hu,Seiberg:2008qj} in this point. 
However, the flavor symmetry $SU(N_F)$ is broken down to $SU(N_F-N_C)\times SU(N_C)$ 
in the SUSY breaking vacua and hence
we have a constraint, $N_F-N_C \ge 5$ or $N_C\ge 5$ for keeping the SM gauge symmetries unbroken. 
From the constraint, $N_C+1\le N_F <\frac{3}{2}N_C$, we obtain $N_C>10$ or $N_C \ge 5$, respectively. 
Because of this large number of $N_C$, we lose the success of the SM gauge coupling unification 
at the GUT scale \cite{Jones:2008ib} for a low-scale gauge mediation~\footnote{If one introduces a multiplet in the  adjoint representation of $SU(N_C)$, 
we can maintain the GUT unification as pointed out in Ref.~\cite{Sato:2009yt}.}. 

The above problem is originated from the constraint on $N_F <\frac{3}{2} N_C$ in the model of Intriligator-Seiberg-Shih (ISS)~\cite{Intriligator:2006dd}, 
as discussed in Ref.~\cite{Kitano:2006xg}. 
However, it has been shown recently~\cite{Izawa:2009nz} that the SUSY is dynamically broken even in the conformal windows of the vector-like theories, that is, for $\frac{3}{2}N_C\le N_F < 3N_C$, if we introduce gauge singlet multiplets.

In this paper, we show that a gauge mediation consistent with the GUT unification is easily constructed in the above conformal windows.
In fact, the gauge mediation dynamics is similar to the one in a semi-direct mediation and
the model possesses the merit of the conformal gauge mediation \cite{Ibe:2007wp}, 
where all the scales of the model are determined only by one parameter, that is, the mass of messengers.
We also discuss a new mechanism for generating SUSY-breaking masses of the gauginos in the SUSY standard model (SSM). 
The generation of the gaugino masses requires some deformation of the model in the ISS model, due to the existence of an R symmetry.
However, we do not need such a deformation in our model.
The SUSY breaking field has a fractional R charge, whose $F$-term breaks the R symmetry as well as the SUSY.
We show that the gauginos acquire the SUSY breaking masses through instanton effects by picking up the $F$-term R breaking.

\section{SUSY breaking in  conformal windows of vector-like gauge theories}
\label{sec:2}
In this section we briefly review the SUSY breaking in  conformal windows of vector-like gauge theories~\cite{Izawa:2009nz}.
We use a slightly different approach from the one in Ref.~\cite{Izawa:2009nz}. The argument of this section is less rigorous 
but may give a more intuitive physical picture for the dynamics of the model.

The model is based on a SUSY $SU(N_C)$ gauge theory with $N_Q$ flavors of quarks $Q^i,~\tilde{Q}_{\tilde i}~~(i,~\tilde{i}=1,\cdots,N_Q)$ 
in the fundamental and anti-fundamental 
representations of $SU(N_C)$, $N_P$ flavors of massive quarks $P^a,~\tilde{P}_a~~(a=1,\cdots,N_P)$ in the same representation as $Q^i,~\tilde{Q}_{\tilde i}$,
and gauge singlet fields $S^{\tilde i}_{~j}$. We omit gauge indices for simplicity. 
We consider the dynamics of the model for $\frac{3}{2}N_C<N_Q+N_P<3N_C$ throughout this paper.
The tree level superpotential of the model is given by
\beq
W=\l \tr(SQ \tilde{Q})+mP\tilde{P} , \label{eq:treesuper}
\eeq
where $\tr(SQ \tilde{Q})=S^{\tilde i}_{~j}Q^j\tilde{Q}_{\tilde i}$ and $P\tilde{P}=P^a\tilde{P}_a$.
In a regime where the mass $m$ can be neglected, this theory has 
an infrared conformal fixed point~\cite{Seiberg:1994pq}
if $\frac{3}{2}N_C<N_Q+N_P<3N_C$ is satisfied. We also assume $N_Q<N_C$ and $N_P>N_C$, as discussed Ref.~in \cite{Izawa:2009nz}. 

Consider the region where the vacuum expectation value (vev) of $S$ is large. Then $Q,~\tilde{Q}$ become massive and we can integrate out them.
After integrating out all quarks, a low-energy gaugino condensation induces an effective superpotential,
(see Ref.~\cite{Intriligator:1995au} for a review)

\beq
W_{\rm eff}=N_C \left(m^{N_P} \L^{3N_C-N_F}\det (\l S)\right)^{1 \o N_C}, \label{eq:effsuper}
\eeq
where $\L$ is the (holomorphic) dynamical scale of the model, and $N_F \equiv N_Q+N_P$. 
One can easily see that the superpotential Eq.~(\ref{eq:effsuper}) is of runaway type. Naively, there seems to be no stable vacuum in the theory.
However, as emphasized in Ref.~\cite{Izawa:2009nz}, 
we must consider the quantum corrections to the K\"ahler potential to determine the behavior of the potential.

To obtain the effective K\"ahler potential, we follow the Wilsonian approach of Ref.~\cite{ArkaniHamed:1997ut}.
Let us consider the effective K\"ahler potential of the fluctuation $\hat{S}^{\tilde i}_{~j}=S^{\tilde i}_{~j}-(S_0)^{\tilde i}_{~j}$
around the background $(S_0)^{\tilde i}_{~j}={\rm const}$. 
We consider the case $(S_0)^{\tilde i}_{~j}=S_0 \d^{\tilde i}_{~i}$ for simplicity.
At energies much higher than the mass of the quarks $Q,\tilde{Q}$ and $P,\tilde{P}$, the K\"ahler potential of $S$ is given by
\beq
K_{\rm eff}=Z_S(M)\tr (S^\dagger S)+\cdots,
\eeq
where $Z_S(M)$ is the wave function renormalization of $S$ at the Wilson cutoff scale $M$, and dots denote higher dimensional operators.
Below the effective mass of $Q,\tilde{Q}$, $m_Q \propto S_0$, the quarks $Q,\tilde{Q}$ decouple from the dynamics at the scale $M$. 
The effective mass $m_Q$ depends on $S_0$, and we consider the region of $S_0$
in which $m_Q$ is much larger than 
the mass of $P,\tilde{P}$.
Then, $S$ has no (relevant or marginal) interactions below the scale $m_Q$. The K\"ahler potential becomes, 
\beq
K_{\rm eff}=Z_S(M)\left[(1+\d_1)\tr \left(\hat{S}^\dagger \hat{S}\right)+\d_2\tr(\hat{S}^\dagger)\tr( \hat{S})\right]+\cdots, \label{eq:effkahler}
\eeq
where $\d_1$ and $\d_2$ are ${\cal O}(1)$ corrections which appear at the threshold $m_Q$.
We neglect $\d_1$ and $\d_2$ in the following discussions since they are not important. 
(For the reader who are interested in more rigorous definition of the effective K\"ahler potential, see Ref.~\cite{Izawa:2009nz}.)

By using Eqs.~(\ref{eq:effsuper}) and (\ref{eq:effkahler}), and setting the fluctuation to be zero, $\hat{S}^{\tilde i}_{~j}=0$, 
the effective potential for $S^{\tilde i}_{~j}=S_0 \d^{\tilde i}_{~j}$ is given by
\beq
V_{\rm eff}(S_0)=Z_S(M)^{-1}N_Q \left| m^{N_P} \L^{3N_C-N_F} \l^{N_Q}  \right|^{2 \o N_C} \left|S_0^2\right|^{-1+\frac{N_Q}{N_C}}.
\eeq
We should take $M\to 0$ to integrate out all momentum modes. The factor $Z_S(M)^{-1}$ represents the effect of the quantum corrections. 

Now we determine $Z_S(M)^{-1}$ for $M \to 0$ as a function of $S_0$.
First, let us determine $m_Q$. 
$m_Q$ is determined by the equation
\beq
m_Q=Z_Q(m_Q)^{-1}|\l S_0|,\label{eq:massdetermin}
\eeq
where $Z_Q(M)$ is the wave function renormalization of $Q,\tilde{Q}$.
$Z_S$ and $Z_Q$ are given by
\beq
Z_S(M)&=&\exp\left(-\int^M_{M_*}\g_S(M') d\log M' \right), \label{eq:eq:swave}\\
Z_Q(M)&=&\exp\left(-\int^M_{M_*}\g_Q(M')d\log M' \right),\label{eq:eq:qwave}
\eeq
where $\g_S$ and $\g_Q$ are the anomalous dimensions of $S$ and $Q,\tilde{Q}$ respectively, 
and $M_*$ is the scale (taken to be larger than all the other scales) at which the fields are normalized 
canonically. 

Notice that the theory is assumed to be on the conformal fixed point above the scale $m_Q$, i.e. $M>m_Q$.
Then, $\g_S$ and $\g_Q$ are constant at the conformal fixed point, which we denote $\g_{S*}$ and $\g_{Q*}$.
They satisfy the relation $\g_{S*}+2\g_{Q*}=0$ which is required by the renormalization
group equation of the Yukawa coupling $\l$ at the fixed point.
Then, we can do the integrations in Eqs.~(\ref{eq:eq:swave}) and (\ref{eq:eq:qwave}) to obtain for $M > m_Q$,
\beq
Z_S(M)=\left(\frac{M}{M_*}\right)^{-\g_{S*}},~~~
Z_Q(M)=\left(\frac{M}{M_*}\right)^{\g_{S*}/2}. \label{eq:wave}
\eeq
Using Eqs.~(\ref{eq:massdetermin}) and (\ref{eq:wave}), we obtain
\beq
m_Q=M_* \left(\frac{|\l S_0|}{M_*}\right)^{1 \o (1+\g_{S*}/2)}.
\eeq

Now let us determine $Z_S(M)$ for $M <m_Q$. Below the effective mass $m_Q$, $\hat{S}$ has no relevant or marginal interaction and hence  
$\g_S(M <m_Q) \simeq 0$. Therefore, $Z_S(M)=Z_S(m_Q)$ for $M < m_Q$ and we obtain
\beq
Z_S(M<m_Q)^{-1}=Z_S(m_Q)^{-1}=\left(\frac{|\l S_0|}{M_*}\right)^{\g_{S*} \o (1+\g_{S*}/2)}. 
\eeq
Thus, the power of $|S_0|$ in the potential is given by,
\beq
V_{\rm eff} &\propto& |S_0^2|^A,
\eeq
where
\beq
A&=&\frac{\g_{S*}/2}{1+\g_{S*}/2}-\left(1-\frac{N_Q}{N_C}\right). \label{eq:beki}
\eeq
If $A$ is positive, the runaway of the potential is stabilized. 
Numerical values of $\g_{S*}$ are listed in Table \ref{table:2}. There exist many sets of $N_C,~N_Q,~N_P$ satisfying the condition $A>0$,
so the runaway can be stopped and the theory has well defined vacua.

\begin{table}[Ht]
\begin{center}
\begin{tabular}{c|c|c|c|c|c|c}
$(N_C,~N_Q~N_P)$&(3,~2,~3)&(3,~2,~4)&(4,~3,~3)&(4,~3,~4)&(4,~3,~5)&(5,~3,~5)                \\ \hline 
$\g_{S*}/2$&0.70&0.36&1&0.59&0.35&0.81
\end{tabular}
\caption{The anomalous dimension of $S$ at the conformal fixed point
for several values of $N_C$, $N_Q$, and $N_P$. This table is taken from Ref.~\cite{Izawa:2009nz}.}
\label{table:2}
\end{center}
\end{table}

The above argument breaks down when $S_0$ becomes small and the effective mass $m_Q$ becomes smaller than the mass of $P,\tilde{P}$.
Therefore, we expect that the potential minimum of the theory exists in the region of small $S_0$.
In Ref.~\cite{Izawa:2009nz}, the SUSY was shown to be broken by an indirect argument using the Witten index~\cite{Witten:1982df}.
In the next section, we will show by using Seiberg duality~\cite{Seiberg:1994pq} that the SUSY is indeed broken at the tree level in the dual magnetic theory.
We will also show explicitly that the potential minimum exists at $\vev{S}=0$ if the theory is very strongly coupled in the electric theory.

\section{Magnetic dual of the theory and SUSY breaking}
\label{sec:3}
In the SUSY breaking model reviewed above, the couplings become very strong after the decoupling of $P^a,~\tilde{P}_a$.
Then, it is convenient to use the Seiberg duality to study the low energy dynamics of the model.
We can see explicitly that the SUSY is broken in the dual theory. Furthermore, as we will see below,
if the couplings are too strong at the conformal fixed point in the electric theory
(so that the dual magnetic theory is quite weakly coupled), the flavor $SU(N_P)$ symmetry 
breaks down spontaneously, implying the $SU(5)_{\rm GUT}$ breaking in the gauge mediation.

\subsection{Seiberg duality}

Before considering the dual of our theory, let us review the original work of Ref.~\cite{Seiberg:1994pq}, 
which motivates the duality of the present model with the singlet $S^{\tilde i}_{~j}$.

Consider an $SU(N_C)$ SUSY QCD with $N_F$ flavors of quarks $Q^i,~{\tilde Q}_{\tilde i}$. If $\frac{3}{2}N_C<N_F<3N_C$, 
this theory flows to a conformal fixed point at low energies. Seiberg argued that this theory is dual to an $SU(N_F-N_C)$ gauge theory
with $N_F$ flavors of quarks $q_i,~{\tilde q}^{\tilde i}$ and $N_F \times N_F$ singlets $M^i_{\tilde j}$, with a superpotential,
\beq
W=\frac{1}{\mu} M^i_{\tilde j} q_i{\tilde q}^{\tilde j},
\eeq
where $\mu$ is related to the electric and magnetic holomorphic dynamical scales $\L$ and ${\tilde \L}$ by~\cite{Intriligator:1995au}
\beq
\L^{3N_C-N_F} {\tilde \L}^{3(N_F-N_C)-N_F} = (-1)^{N_F-N_C}\mu^{N_F}, \label{eq:matching}
\eeq
and the singlets $M^i_{\tilde j}$ are the mesons $M^i_{\tilde j}=Q^i {\tilde Q}_{\tilde j}$ of the electric theory. 

Let us take a dual of the dual. The dual of the dual theory is an $SU(N_C)$ gauge theory with quarks 
$Q'^i,~{\tilde Q'}_{\tilde j}$, and singlets $M_i^{\tilde j}$, $N_i^{\tilde j} \equiv q_i{\tilde q}^{\tilde j}$ with a superpotential
\beq
W&=&\frac{1}{\mu} M^i_{\tilde j} q_i{\tilde q}^{\tilde j}+\frac{1}{{\tilde{\mu}}}N_i^{\tilde j}Q'^i{\tilde Q'}_{\tilde j} \nonumber \\
&=&\frac{1}{\mu} (M^i_{\tilde j}-Q'^i {\tilde Q'}_{\tilde j})N_i^{\tilde j},
\eeq   
where ${\tilde \mu}=-\mu$. From this superpotential, $M$ and $N$ become massive 
and we can integrate out these fields. Then we obtain $N_i^{\tilde j}=0$ and 
$M^i_{\tilde j}=Q'^i {\tilde Q'}_{\tilde j}$, recovering the original electric theory, provided with $Q'^i=Q^i$ and $\tilde{Q}'_{\tilde i}=\tilde{Q}_{\tilde i}$.

Now let us apply the above considerations to our model with $m=0$.
First, we define mesons~\footnote{
When $m=0$, there are $SU(N_P)\times SU(N_P)$ symmetry acting on $P$ and $\tilde{P}$ separately.
Thus we use different indices $a$ and $\tilde{a}$ for them.} 
$K^a_{~\tilde b}=P^a {\tilde P}_{\tilde b}$, $L^a_{~\tilde i}=P^a {\tilde Q}_{\tilde i}$, $\tilde{L}^i_{~\tilde a}=Q^i {\tilde P}_{\tilde a}$ 
and $N^i_{~\tilde j}=Q^i {\tilde Q}_{\tilde j}$. Second, consider an $SU(N_F-N_C)$ gauge theory ($N_F=N_Q+N_P$) with quarks
$q_i$, ${\tilde q}^{\tilde j}$, $p_a$ and ${\tilde p}^{\tilde b}$ in the fundamental and anti-fundamental representation of $SU(N_F-N_C)$.
The superpotential is, 
\beq
W=\l \tr (SN)  + \frac{1}{\mu} \left\{ \tr(N\tilde{q}q)+\tr(L\tilde{q}p)+\tr(\tilde{L}\tilde{p}q)+\tr(K\tilde{p}p)   \right\},
\eeq
where contractions of flavor indices are represented by trace as before.
The first term in this superpotential is the one present in the original electric theory,
and other terms appear because of taking duality. 
From this superpotential, we can see that $S$ and $N$ become massive, as in the dual of the dual theory considered above.
Thus, we can integrate them out, and obtain $N=0$, $\l S = -\frac{1}{\mu}\tilde{q}q$.
Then we finally obtain a theory with mesons $K$, $L$, $\tilde{L}$, quarks $q$, $\tilde{q}$, $p$, $\tilde{p}$ and a superpotential,
\beq
W=\frac{1}{\mu} \left\{\tr(L\tilde{q}p)+\tr(\tilde{L}\tilde{p}q)+\tr(K\tilde{p}p)   \right\}.
\eeq 
This is the dual of our theory with $m=0$. We believe that this duality is correct because it is a straightforward extension of the original Seiberg's duality.

As a check, let us consider the 't~Hooft anomaly matching condition. We see, in the following argument, that the anomaly matching condition is indeed satisfied. 
Without the singlet $S$, the duality is the original one considered in Ref.~\cite{Seiberg:1994pq}, and the anomaly matching condition is satisfied.
Then, introducing $S$ in both the electric theory and the magnetic theory, the global symmetry of the theory reduces to a subgroup of the one
in the original theory.
The anomaly is still matched, since we have only added the same singlet $S$ to both the electric and magnetic theory.
Finally, let us integrate out $S$ and $N$ in the magnetic theory.
Massive fields in general do not contribute to anomalies, so it has no effect on the anomaly matching to integrate out
the massive fields $N$ and $S$ in the magnetic theory. Thus, we can conclude that the anomaly matching condition is satisfied in our duality.

\subsection{Low energy dynamics}
In this subsection, we analyze the dual magnetic theory of the present model with $m \neq 0$. 
We take a more general mass term $\tr(mP\tilde{P})=m^{\tilde b}_{~a}P^a\tilde{P}_{\tilde b}$.
Then we have an additional term in the superpotential, $\tr(mK)$, in the dual theory (see Eq.~(\ref{eq:dualsuperpot})). 

Although the couplings of the model are uniquely determined by $N_C,~N_Q$ and $N_P$  at the fixed point,
we can make the dual theory very weakly coupled as follows. 
Consider a parameter region 
where masses have a hierarchy; $m=\textrm{diag}(m_1,~m_2,~\cdots,~m_{N_P})$ with
$|m_{N_P}| \gg \cdots \gg |m_1|$. In this case, we can integrate out massive quarks $P$, $\tilde{P}$ step by step in the electric theory,
and eventually the electric theory enters into confining phase. In the magnetic theory,
the mass $m_{N_P}$ induces a vacuum expectation value for $\tilde{p}_{N_P} p^{N_P}$, and hence gauge symmetry is broken down to $SU(N_F-N_C-1)$.
Then, some fields become massive and we obtain the theory with $N_F-N_C \to N_F-N_C-1$ and $N_P \to N_P-1$.
This process can be continued and eventually we obtain an asymptotic non-free theory in the magnetic description. 
In this way, we reach an asymptotic non-free theory where a weak coupling analysis becomes reliable. In the following analyses, 
we assume that the dual magnetic theory is weakly coupled (i.e. we assume the mass hierarchy discussed above or a very weak coupling at the fixed point
in the magnetic theory).

The tree level superpotential is now given by
\beq
W_{\rm tree} = \frac{1}{\mu} \left\{\tr(L\tilde{q}p)+\tr(\tilde{L}\tilde{p}q)+\tr(K\tilde{p}p)   \right\}+\tr (mK). \label{eq:dualsuperpot}
\eeq
From this superpotential, we can see the SUSY breaking very easily. $F$-term of $K$ is
\beq
-F_K^\dagger \propto \frac{\q W_{\rm tree}}{\q K}=\frac{1}{\mu}\tilde{p}p + m. \label{eq:Fcondition}
\eeq 
This equation is an $N_P \times N_P$ matrix equation, and  $\tilde{p}$ and $p$ are $N_P \times (N_F-N_C)$ and 
$(N_F-N_C) \times N_P$ matrices, respectively. Because we have assumed $N_Q<N_C$ to obtain a runaway superpotential in the electric theory, we have
$(N_F-N_C)-N_P=N_Q-N_C<0$, so $\textrm{rank}(p)<N_P$.  Thus we can conclude that Eq.~(\ref{eq:Fcondition}) cannot be zero
and the SUSY is broken, since $\textrm{rank}(\tilde{p}p)<N_P$ and $\textrm{rank}(m)=N_P$.
This is the ``rank condition SUSY breaking'' as in the Intriligator-Seiberg-Shih (ISS) model~\cite{Intriligator:2006dd}. 
If $N_Q \geq N_C $, Eq.~(\ref{eq:Fcondition}) can be zero
and the SUSY is not broken. This is consistent because we know that in this case ($N_Q \geq N_C$) there are no runaway superpotentials and 
SUSY vacua exist in the electric theory. It is remarkable that the rank condition for the SUSY breaking in the magnetic theory coincides with the runaway
condition in the electric theory.

What happens if we take into account non-perturbative effects? 
It is known that a dynamically generated superpotential restores the SUSY in the ISS model~\cite{Intriligator:2006dd}.
However, in the present model, the SUSY is broken even if non-perturbative effects are taken into account.
Suppose that mesons $K$, $L$, and $\tilde{L}$ have vevs and all the quarks become massive. (This is possible only if $N_Q \leq N_P$. If
$N_Q>N_P$, some quarks have to be massless. Note that we have assumed $N_Q<N_C<N_P$ in the previous section.)  
Then the following superpotential is generated by gaugino condensation,
\beq
W_{\rm dyn}&=&(N_F-N_C)\left( \tilde{ \L}^{3(N_F-N_C)-N_F}  \det (M/\mu)  \right)^{\frac{1}{N_F-N_C}} \nonumber \\
&=&-(N_F-N_C)\left(  \L^{-(3N_C-N_F)}  \det M  \right)^{\frac{1}{N_F-N_C}},\label{eq:nonpsuper}
\eeq
where we have used Eq.~(\ref{eq:matching}), and $M$ is defined by
\beq
M=
\left(
\begin{array}{cc}
K&L \\
\tilde{L}&0
\end{array}
\right).
\eeq
 Notice that $M^a_{~\tilde b}=K^a_{~\tilde b}$, $M^a_{~\tilde i}=L^a_{~\tilde i}$, $M^i_{~\tilde a}=\tilde{L}^i_{~\tilde a}$ 
 and $M^i_{~\tilde j}=0$. 
 
 Integrating out the massive quarks, the superpotential is 
\beq
W_{\rm eff}=\tr (mK) -(N_F-N_C)\left(  \L^{-(3N_C-N_F)}  \det M  \right)^{\frac{1}{N_F-N_C}}.
\eeq
The $F$-term of $K$ is 
\beq
-(F^\dagger_K)^{\tilde a}_{~b} \propto \frac{\q W}{\q K^b_{~\tilde a}}= m^{\tilde a}_{~b}-\left(  \L^{-(3N_C-N_F)}  \det M  \right)^{\frac{1}{N_F-N_C}}
(M^{-1})^{\tilde a}_{~b}.
\eeq
For this $F$-term to vanish, the inverse matrix $M^{-1}$ must be of the form
\beq
M^{-1}=\left(
\begin{array}{cc}
\a m & A \\
B & C
\end{array}
\right),
\eeq
where $\a$ is some non-zero constant and $A,~B$ and $C$ are some matrices. However, $M^{-1}$ of the above form does not exist.
The product of the above $M^{-1}$ and $M$ is
\beq
\left(
\begin{array}{cc}
1&0 \\
0&1
\end{array}
\right)=
M^{-1}M=\left(
\begin{array}{cc}
\a mK+A\tilde{L} & \a mL \\
BK+C\tilde{L} & BL
\end{array}
\right).
\eeq
The Equation $\a mL=0$ implies $L=0$ because $\a$ is non-zero and $m$ is an invertible matrix, but this contradicts 
the equation $BL=1$. Thus the $F$-term of $K$ cannot vanish. This shows that even if the non-perturbative effect Eq.~(\ref{eq:nonpsuper}) is taken into account, 
SUSY cannot be restored. We see that $M^i_{~\tilde j}=0$ plays a crucial role in the above proof which comes from the integration of the singlet $S^{\tilde i}_{~j}$.

Finally, we show another evidence supporting the duality in the present model.
In the above consideration, we have investigated the direction in the classical moduli space 
where $K,~L$ and $\tilde{L}$ have vevs and all the quarks become massive, so that the quarks have 
vanishing vevs. Now we consider another direction in the classical moduli space.
In the case $m=0$, 
the vevs of $\tilde{p}p,~\tilde{q}p$ and $\tilde{p}q$ are constrained to be zero by the equations of motion
of $K,~L$ and $\tilde{L}$, respectively, but the vevs of $\tilde{q}q$ are not. Let us consider the case that 
$K$ and $\tilde{q}q$
have very large vevs. These vevs give masses to $p,~\tilde{p},~L$ and $\tilde{L}$, and the Affleck-Dine-Seiberg superpotential~\cite{Affleck:1983mk}
is generated at low energies with a dynamical scale $\tilde{\L}_L^{3(N_F-N_C)-N_Q}=\tilde{\L}^{3(N_F-N_C)-N_F}\det(K/\mu)$ as
\beq
W_{\rm dyn}&=&(N_F-N_C-N_Q)\left( \frac{\tilde{\L}_L^{3(N_F-N_C)-N_Q}}{\det \tilde{q}q} \right)^{\frac{1}{N_F-N_C-N_Q}} \nonumber \\
&=&(N_P-N_C)\left( \frac{\tilde{\L}^{3(N_F-N_C)-N_F}\det(K/\mu)}{\det \tilde{q}q} \right)^{\frac{1}{N_P-N_C}}.
\eeq
Adding the mass term $\tr(mK)$ to this superpotential, solving the equation of motion of $K$, and using Eq.~(\ref{eq:matching}), we obtain
\beq
K=m^{-1}\left(\L^{3N_C-N_F} \det m \det \left( \frac{\tilde{q}q}{-\mu} \right)  \right)^{\frac{1}{N_C}},
\eeq
and an effective superpotential,
\beq
W_{\rm eff}=N_C \left(\L^{3N_C-N_F} \det m \det \left( \frac{\tilde{q}q}{-\mu} \right)  \right)^{\frac{1}{N_C}}. \label{eq:effsuper2mag}
\eeq
In the process of taking duality, we have found that the electric variable $S$ and the magnetic variables $q,~\tilde{q}$
are related by $\l S=-\frac{1}{\mu}\tilde{q}q$. Using this relation, Eq.~(\ref{eq:effsuper2mag}) is just the superpotential Eq.~(\ref{eq:effsuper}) derived in the
electric theory. This is another evidence for the correctness of our duality.

\subsection{GUT breaking and R symmetry breaking} \label{sec:gutandrbreak}
For a gauge mediation model to be phenomenologically viable, following conditions must be satisfied;
\begin{enumerate}
\item The standard model gauge group $SU(3)\times SU(2)\times U(1)$ should not be broken spontaneously by the SUSY breaking dynamics.
\item R symmetry (if exist) should be broken to generate the gaugino masses.
\end{enumerate}
Unfortunately, both of the conditions may {\it not} be satisfied if the dual theory is weakly coupled.
To see this, consider Eq.~(\ref{eq:Fcondition}) with $m^{\tilde a}_{~b}=m\d^{\tilde a}_{~b}$.
From Eq.~(\ref{eq:Fcondition}), we see that $p$ develops a vev of the form
\beq
p=\left(
\begin{array}{cc}
\sqrt{-\m m}\cdot {\bf 1}_{(N_F-N_C)\times (N_F-N_C)} ~,& {\bf 0}_{(N_F-N_C)\times (N_C-N_Q)}
\end{array}\right),\label{eq:pvev}
\eeq
and similarly for $\tilde{p}$. This vev breaks the flavor $SU(N_P)$ symmetry. 
Because we want to identify a (sub)group of $SU(N_P)$ with the GUT gauge group $SU(5)_{\rm GUT}$, the SM gauge group may be broken down.
Of course it is possible that the SM gauge group is in a residual symmetry group after the breaking of $SU(N_P)$ 
as in the case of direct mediation models in the ISS model, but in that case the Landau pole problem of the SM gauge couplings is 
unavoidable for the low-scale gauge mediation.
Furthermore, there is an R symmetry with the charge assignment,
\beq
Q,\tilde{Q}~:~1-\frac{N_C}{N_Q},~~~~~~~P,\tilde{P}~:~1,~~~~~~~S~:~\frac{2N_C}{N_Q},\label{eq:Rcharge1}
\eeq
or equivalently,
\beq
q,\tilde{q}~:~\frac{N_C}{N_Q},~~~~~~~p,\tilde{p}~:~0,~~~~~~~K~:~2,~~~~~~~L,\tilde{L}~:~2-\frac{N_C}{N_Q}.
\eeq
The vev Eq.~(\ref{eq:pvev}) does not break this R symmetry, so the gaugino masses in the SSM are not generated.

The breaking of $SU(5)_{\rm GUT}$ and the non-breaking of $U(1)_R$ should be regarded as a consequence of weak couplings in the dual magnetic theory.
Consider the other limit, i.e. weak couplings at the fixed point in the electric theory. 
Then, it is convenient to use the electric description of the dynamics. 
In this case, the gauge and Yakawa couplings above the mass threshold of $P,\tilde{P}$ are weak and hence the low energy
dynamical scale (at which SUSY is broken) is much lower than the mass of $P,~\tilde{P}$. Then $P,~\tilde{P}$ are decoupled at the low energies much before 
the couplings become strong, so it is highly unlikely that those fields, $P$ and $\tilde{P}$, develop vevs. Therefore, 
it is very natural to assume that the $SU(N_P)$ symmetry remains unbroken. 
Furthermore, $F_K=F_{P\tilde{P}}$ is likely to be zero, so the SUSY breaking must be developed by other fields.
Here recall that the SUSY is broken in the present model as shown in Ref.~\cite{Izawa:2009nz}.
After the decoupling of $P,\tilde{P}$, the gauge invariant chiral fields at the low energies 
are $S$ and $Q\tilde{Q}$. Then, it is reasonable to consider that at least one of those fields develop $F$ terms~\footnote{
If the equation of motion of the chiral field $S$, $\frac{1}{4}\bar{D}^2S^\dagger=\l Q\tilde{Q}$, 
is correct as an operator equation, then by taking the components we obtain $-\vev{F_S}^\dagger=\l \vev{Q\tilde{Q}}$ 
and $-\q^\mu\q_\mu \vev{S}^\dagger=\l \vev{F_{Q\tilde{Q}}}$,
where the lowest components of the chiral fields are denoted by the same symbol as the chiral fields themselves.
$\q^\mu\q_\mu \vev{S}^\dagger$ vanishes because Lorentz invariance is not broken, so we have $\vev{F_{Q\tilde{Q}}}=0$. 
Thus the SUSY breaking is perhaps induced by 
$\vev{F_S}$. However, we cannot exclude a possibility that $\vev{F_S}$ is also zero, 
and the SUSY breaking is induced by a vev of some other vector superfield operator.}.
Now, it is important that the $F$ terms of those fields carry nonzero $U(1)_R$ charges, so the R symmetry is also broken by the $F$ terms. 
For example, the $F$ term of $S$, $F_S$, 
has $R$ charge $-2(1-N_C/N_Q)$ as seen from Eq.~(\ref{eq:Rcharge1}), which may be useful for a gauge mediation as we will see in the next section.

The above consideration suggests that a phase transition occurs as the strengths of couplings are changed. 
We have seen that $F_S \sim F_{q\tilde q}=0$ in the tree level analyses in the dual magnetic description when the coupling is too strong in the electric theory. 
However, when the coupling is weak at the fixed point in the electric theory
so that the dynamical scale is much smaller than the physical mass of $P,\tilde{P}$, the $S$ and $Q$ most likely have non-vanishing $F$-terms.
We assume that $F_S \neq 0$ in the next section. See \ref{app:A} for an explicit toy model where a similar phase transition from $F_S =0$ to $F_S \neq 0$ occurs.

\section{Gauge mediation model}
\label{sec:4}
Now let us consider a candidate of the semi-direct gauge mediation model with the above SUSY breaking mechanism.
We identify a subgroup of the flavor $SU(N_P)$ symmetry as the $SU(5)_{\rm GUT}$ gauge group.
Our model is an explicit example of the strongly coupled semi-direct gauge mediation of Ref.~\cite{Izawa:2008ef}.
We impose the following conditions on $(N_C,~N_Q,~N_P)$.
\begin{enumerate}
\item $N_C \leq 4$. From the point of view of $SU(5)_{\rm GUT}$ gauge group, the color number $N_C$ of 
the hidden gauge group becomes the messenger number
in gauge mediation. For the perturbative GUT unification to be maintained, the messenger number must be small. In particular, 
for the low-scale gauge mediation, the constraint on the messenger number is rather severe, $N_C \leq 5$ with $N_C=5$ marginal~\cite{Jones:2008ib}.
In our case, the messenger fields have large negative anomalous dimension, 
which effectively increase the messenger number in the SM $\b$ functions~\cite{Sato:2009yt}. 
Thus we impose $N_C\leq4$ in this paper~\footnote{In fact, a large negative anomalous dimension of $P,\tilde{P}$ makes the messenger contribution
to the SM $\b$ function effectively larger than 5 even for $N_C=4$. 
So we should assume that the theory is off the conformal fixed point and the couplings are small at high energies.
}. (However, see Ref.~\cite{Sato:2009yt} for a mechanism which allows $N_C \geq 5$ without spoilng 
the perturbative GUT unification.)
\item $N_P=5$. To identify a subgroup of the flavor $SU(N_P)$ symmetry with $SU(5)_{\rm GUT}$, we only need $N_P \geq 5$ as a 
necessary condition. However, it is more appealing to take $N_P=5$, because when $N_P>5$, there seems to be no reason for the mass of messenger 
quarks and the other $N_P-5$ flavors of quarks to be the same. Thus, to achieve a one parameter model of SUSY breaking and gauge mediation, that is,
the conformal gauge mediation~\cite{Ibe:2007wp}, it is more desirable to take $N_P=5$.
\item $N_Q<N_C$ and $A>0$ in Eq.~(\ref{eq:beki}). This is the requirement for the present SUSY breaking dynamics to work. 
\end{enumerate}
Imposition of those conditions uniquely leads to the model $(N_C,~N_Q,~N_P)=(4,~3,~5)$. We call this model the $SU(4)$ model from now on.

In the $SU(4)$ model, there is a relation $N_F=2N_C$, i.e. in the middle of conformal window. 
This relation means that the model is strongly coupled at the fixed point 
in both the electric description and the magnetic description. The strong coupling in the electric description means that
the dynamical scale $\L_{\rm phys}$ and the mass of $P,\tilde{P}$, $m_{\rm phys}$ are of the same order,
 $\L_{\rm phys} \sim m_{\rm phys}$. This is desirable~\cite{Ibe:2007wp} because we may obtain the same order of gaugino and sfermion masses, 
and also obtain a light gravitino mass $m_{3/2}\lsim {\mathcal O}(10)~\EV$, which is free from the cosmological gravitino problems~\cite{Viel:2005qj}. 
The strong coupling in the dual magnetic description means that
$SU(5)_{\rm GUT}$ breaking and $U(1)_R$ non-breaking argument discussed in the previous section is not applicable to this model.
Because the electric theory is also strongly coupled, we can say nothing about the spontaneous breaking of those symmetries.
In the following, we assume that $SU(5)_{\rm GUT}$ is not broken down, and also assume $F_S \neq 0$ as discussed in the previous section.
Notice that the SUSY is broken whether the theory is strongly coupled or not, as shown in Ref.~\cite{Izawa:2009nz}. 

Let us investigate the dynamics of the model at the fixed point at the 1-loop level. The 1-loop anomalous dimensions of $S,~Q,$ and $P$ are given by
\beq
\g_S^{\rm 1-loop} = N_C\frac{ |\l|^2}{8\pi^2},~~~~\g_Q^{\rm 1-loop}=N_Q\frac{|\l|^2}{8\pi^2} - \frac{N_C^2-1}{N_C} \frac{g^2}{8\pi^2},~~~~\g_P^{\rm 1-loop}= 
- \frac{N_C^2-1}{N_C} \frac{g^2}{8\pi^2}.
\eeq
The RG running equations of the gauge and Yukawa couplings are given by
\beq
 M\frac{d }{d M }|\l|^2 &=& (\g_S+2\g_Q)|\l|^2, \label{eq:lbeta}  \\
 M\frac{d }{d M }g^2  &=& -\frac{g^4}{8\pi^2} \frac{3N_C-(1-\g_Q)N_Q-(1-\g_P)N_P}{1-N_Cg^2/8\pi^2} \label{eq:gbeta},
\eeq
where we have adopted the NSVZ $\beta$ function~\cite{Novikov:1983uc} for the gauge coupling $\b$ function. Requiring these $\beta$ functions to vanish and
substituting $(N_C,~N_Q,~N_P)=(4,~3,~5)$, we obtain the fixed point values of the coupling constants as
\beq
\left. \frac{|\l_*|^2}{8\pi^2}\right|_{\rm 1-loop}=\frac{4}{31}\simeq 0.13,~~~~~\left. \frac{g_*^2}{8\pi^2}\right|_{\rm 1-loop}=\frac{16}{93}\simeq 0.17,
\eeq
and
\beq
\g_S^{\rm 1-loop} = \frac{16}{31}\simeq 0.52,~~~\g_Q^{\rm 1-loop}=-\frac{8}{31}\simeq -0.26,  ~~~\g_P^{\rm 1-loop}= 
- \frac{20}{31}\simeq -0.65.
\eeq
The exact values of the anomalous dimensions determined by the $a$-maximization technique~\cite{Intriligator:2003jj} are given~\cite{Izawa:2009nz} by
\beq
\g_S \simeq 0.70,~~~~~~~~~~~\g_Q \simeq -0.35,  ~~~~~~~~~~~\g_P \simeq -0.59.
\eeq
The difference between the 1-loop and exact values indicates that the fixed point theory is strongly coupled. 
Although the 1-loop approximation is not a good one, if we evaluate the dynamical scale $\L_{\rm phys}$ from a simple matching at the 1-loop level, we obtain
\beq
 \L_{\rm phys} \sim \exp\left(-\frac{8\pi^2}{(3N_C-N_Q)g_*^2}\right)m_{\rm phys} \sim 0.52 \times m_{\rm phys}.
\eeq 
Thus we can expect that $\L_{\rm phys}$ and $m_{\rm phys}$ are almost the same order. 

Examples of operators generating the sfermion and gaugino masses are as follows~\cite{Ibe:2007wp}.
The lowest dimensional operator which generates the sfermion masses are given by
\beq
 \int d^4\h \left(\frac{g^2_{\rm SM}}{16\pi^2}\right)^2 \frac{-c_1}{m_{\rm phys}^2}\tr(S^\dagger S) \f^\dagger \f,
\eeq
where $\f$ is an SSM field, $g_{\rm SM}$ the SM couplings, and $c_1$ an ${\cal O}(1)$ constant. We assume that $c_1$ is positive~\footnote{
If $c_1$ is negative, we may gauge the $SU(N_P)$ symmetry by another gauge group (not by $SU(5)_{\rm GUT}$), and also introduce a vector-like massive
pair of quarks in the bifundamental representation of $SU(N_P) \times SU(5)_{\rm GUT}$. 
Then, we may have positive sfermion masses due to the property of the semi-direct gauge mediation discussed in Ref.~\cite{Argurio:2009ge}.} 
as in Ref.~\cite{Ibe:2007wp}.
At the tree level, this term generates the sfermion masses of order
\beq
m^2_{\rm sfermion} \sim c_1\left(\frac{g^2_{\rm SM}}{16\pi^2}\right)^2 \frac{|F_S|^2}{m_{\rm phys}^2},
\eeq
with $F_S$ of order $\L_{\rm phys} \sim m_{\rm phys}$. 

The lowest dimensional operator generating the SSM gaugino masses (which respect the R symmetry) is given by
\beq
\int d^4 \h \frac{c_2}{m_{\rm phys}^6}\left(\frac{1}{16\pi^2}\right) \tr (S^\dagger S S^\dagger D^2 S) W_{\rm SM}^2,
\eeq
where $W_{\rm SM}$ is the field strength chiral field of the SSM gauge field with the kinetic term normalized as 
$\int d^2\h (1/4g^2_{\rm SM})W_{\rm SM}^2+{\rm h.c.}$ This term generates the gaugino masses of order
\beq
m_{\rm gaugino} \sim c_2 \left(\frac{g^2_{\rm SM}}{16\pi^2}\right) \frac{|F_S|^2\vev{S}^\dagger F_S}{m_{\rm phys}^6}.
\eeq
This is nonzero only if the vev of $S$, $\vev{S}$, is nonzero. However, there may be no convincing argument to show $\vev{S} \neq 0$ and hence it is desirable 
to find another mechanism for the gaugino mass generation.

Fortunately, there is an operator which generates the gaugino masses even if $\vev{S}=0$. 
Let us consider an $SU(N_C)$ theory with the flavor number of $Q,\tilde{Q}$ given by $N_Q=N_C-1$ (in the $SU(4)$ model, $N_C=4$ and $N_Q=3$).
Then, the R charge of $S$ is given by $2N_C/N_Q=2+2/N_Q$,
and the R charge of $F_S$ is $2/N_Q$. Thus, the nonzero vev for $F_S$ breaks the R symmetry spontaneously~\footnote{
This is a notable difference from the case of the $Sp(2)$ model considered in Ref.~\cite{Ibe:2007wp}. In that model, the R charge of the SUSY breaking
field $S$ is 2, so it is necessary to have $\vev{S} \neq 0$ as well as $F_S \neq 0$ for the generation of the gaugino masses.}.
In fact, there is an operator which respect the R symmetry,
\beq
\int d^4 \h c_3 \left(\frac{1}{16\pi^2}\right) \frac{ (\L_L^\dagger)^{2N_C+1}}{m_{\rm phys}^{4N_C+2}} \tr(S^\dagger S) \det(\bar{D}^2S^\dagger) W_{\rm SM}^2, \label{eq:1instop}
\eeq
where $\L_L$ is the holomorphic dynamical scale below the threshold of $P,\tilde{P}$ (see the discussion below).
Notice that the operator $\bar{D}^2 S^\dagger = -4F_S^\dagger +\cdots$ has R charge $-2/N_Q$, so the operator $\det(\bar{D}^2S^\dagger)$ has R charge $-2$. 
This operator gives the gaugino masses of order
\beq
m_{\rm gaugino} \sim c_3 \left(\frac{g^2_{\rm SM}}{16\pi^2}\right) \frac{(\L_L^\dagger)^{2N_C+1} |F_S|^2 (F_S^\dagger)^{N_Q}}{m_{\rm phys}^{4N_C+2}}. \label{eq:gauginomass2}
\eeq

One can see that Eq.~(\ref{eq:1instop}) may be generated by one-anti-instanton effect, by considering an anomalous $U(1)$ symmetry 
which we call $U(1)_{A}$.
Assign the $U(1)_{A}$ charge for the fields as $Q,\tilde{Q}:+1$, $S:-2$ and $P,\tilde{P}:0$. Then, this is a symmetry at the classical level. 
However, the $U(1)_A$ transformation $Q \to e^{i\a}Q$ and $\tilde{Q} \to e^{i\a}\tilde{Q}$ have the $SU(N_C)$ anomaly, inducing the shift of Lagrangian as 
\beq
\d {\cal L}=\frac{2N_Q\a}{32\pi^2}F_{\mu\nu}\tilde{F}^{\mu\nu},
\eeq
where $F_{\mu\nu}$ is the field strength of $SU(N_C)$ and $\tilde{F}_{\mu\nu}=\frac{1}{2}\e_{\mu\nu\r\s}F^{\r\s}$. This anomaly can be cancelled by
the sift of the vacuum angle of $SU(N_C)$, $\h \to \h+2N_Q\a$, where the topological term in the Lagrangian is given by
\beq
{\cal L}_{\h}=-\frac{\h}{32\pi^2}F_{\mu\nu}\tilde{F}^{\mu\nu}.
\eeq
Therefore, $U(1)_A$ becomes a symmetry if we assign $U(1)_A$ charge $2N_Q$ to $\exp(i\h)$. 
Because $\det(\bar{D}^2S^\dagger)$ has $U(1)_{A}$ charge $2N_Q$,
the $U(1)_{A}$ charge of $\exp(i\h)$ implies that  the operator Eq.~(\ref{eq:1instop}) must be accompanied with the anti-instanton factor,
\beq
(\L_L^\dagger)^{2N_C+1} \equiv M^{2N_C+1}\exp(-S_{\rm anti-inst})= M^{2N_C+1} \exp\left(-8\pi^2/g^2(M)-i\h \right),
\eeq
where $M$ is a renormalization scale below the threshold of $P,\tilde{P}$, and $S_{\rm anti-inst}=8\pi^2/g^2+i\h$ is the anti-instanton classical action.
$(\L_L^\dagger)^{2N_C+1}$ has $U(1)_A$ charge $-2N_Q$. This is the reason for the appearance of $(\L_L^\dagger)^{2N_C+1}$ in Eq.~(\ref{eq:1instop}).

\begin{figure}
\begin{center}
\includegraphics[scale=0.5]{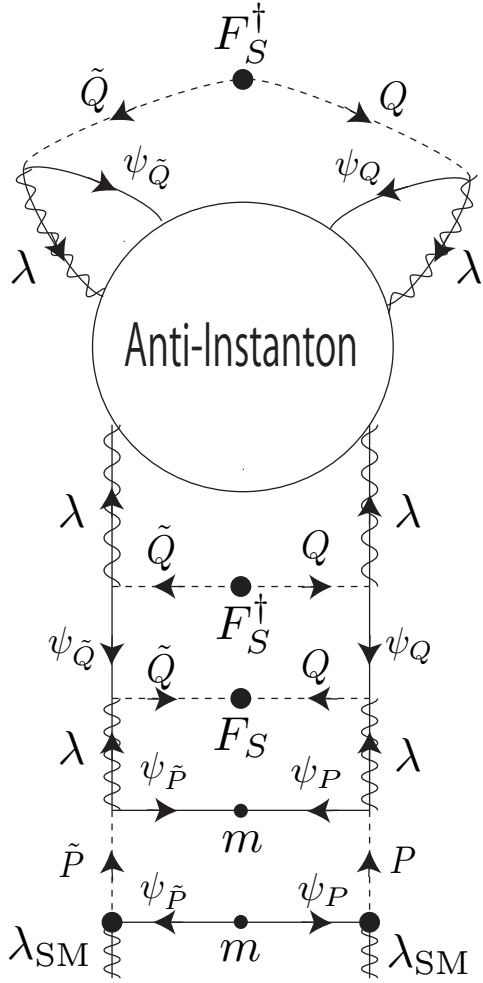}
\caption{An anti-instanton diagram which may generate the operator (\ref{eq:1instop}). We take $N_C=N_Q+1=2$, for simplicity. 
$\l$ and $\l_{\rm SM}$ are the gauginos in the $SU(N_C)$ and the SM gauge theory, respectively. 
$\p_Q,\p_{\tilde Q},\p_{P}$ and $\p_{\tilde P}$ are the fermionic components of the chiral fields $Q,\tilde{Q},P,$ and $\tilde{P}$,
respectively. The scalar components of the chiral fields are denoted by the same symbol as the chiral fields itself.
There are $2N_C$ zero modes for $\l^\dagger$, and one zero mode for each $\p_Q^\dagger, \p_{\tilde Q}^\dagger,\p_P^\dagger$, and $\p_{\tilde P}^\dagger$ 
in the anti-instanton background ($\p$'s are right-handed and $\p^\dagger$'s are left-handed).
The zero modes of $\p_{P},\p_{\tilde{P}}$ are contracted by their mass term $m\p_{P} \p_{\tilde{P}}$, and are not written in the diagram. 
The appearance of $\L_L^{2N_C+1}=m^{N_P}\L^{2N_C+1-N_P}$ in Eq.~(\ref{eq:1instop}) instead of $\L^{2N_C+1-N_P}$ (where $\L^{2N_C+1-N_P}$ 
is the dynamical scale defined above the 
threshold of $P,\tilde{P}$) is due to this contraction of zero
modes of $\p_{P},\p_{\tilde{P}}$ by the mass term. 
The diagram may generate an effective Lagrangian ${\cal L} \propto F_S^\dagger |F_S|^2 \l_{\rm SM}\l_{\rm SM}$.}
\label{fig:1}
\end{center}
\end{figure}

In Fig.~\ref{fig:1}, we show an example of an anti-instanton diagram which may generate the operator Eq.~(\ref{eq:1instop}).
We have not done the computation of diagrams explicitly, and hence Fig.~\ref{fig:1} should be taken only as a schematic picture. 
The existence of a diagram does not always mean the existence of a non-vanishing operator in SUSY theory, because there is a possibility 
that cancellation may occur among various diagrams~\footnote{See also the footnote 6 in the first paper of Ref.~\cite{Ibe:2007wp}}. 
But the operator Eq.~(\ref{eq:1instop}) respects all the symmetry of the theory
and it is not protected by holomorphy, so there seems to be no mechanism which forbids the existence of the operator.
It seems, at the first glance, that Eq.~(\ref{eq:1instop}) is quite small since it is generated by the anti-instanton effect. 
However, the contribution Eq.~(\ref{eq:gauginomass2}) may give not-so-small gaugino masses compared to the sfermion masses, 
since the gauge coupling is large.

\section*{Acknowledgements}
We would like to thank K.-I.~Izawa and F.~Takahashi for the collaboration in the early project, and 
E.~Nakamura for useful discussions on the GUT breaking in semi-direct-type gauge mediation.
This work was supported by 
World Premier International Research Center Initiative
(WPI Initiative), MEXT, Japan.
The work of KY is supported 
in part by JSPS Research Fellowships for Young Scientists.

\appendix
\setcounter{equation}{0}
\renewcommand{\theequation}{\Alph{section}.\arabic{equation}}
\renewcommand{\thesection}{Appendix~\Alph{section}}

\section{Example of phase transition} \label{app:A}
\setcounter{equation}{0}

To see an example of the phase transition discussed in Section~\ref{sec:gutandrbreak}, let us consider a toy model
which have a similarity to our model.  

Consider an $SU(2)$ IYIT model of SUSY breaking \cite{Izawa:1996pk,Intriligator:1996pu}, with one extra massive flavor~\footnote{
This toy model is a simplified version of the model studied by E.~Nakamura~\cite{Nakamura}. 
We thank him for explanation of his result.}. 
The matter chiral fields of the model are quarks
$Q^i,~P^a~~(i=1,2,3,4~~a=1,2)$ in the fundamental representation of the $SU(2)$ gauge group, and six singlets $S_{ij}=-S_{ji}$. 
We take the tree level superpotential to be
\beq
W_{\rm tree}=\frac{1}{2}\l S_{ij}Q^iQ^j + mP^1P^2.
\eeq
One can easily see an analogy between this toy model and our model, although there is neither 
a runaway superpotential nor a conformal fixed point in this toy model.  

Let us consider two limits of this toy model; $m \gg \L$ and $m \ll \L$, where $\L$ is the dynamical scale of 
the gauge theory. In the following analysis, we neglect any perturbative effects and RG evolution of parameters, and 
only consider the strong gauge dynamics.

First, consider the limit $m \gg \L$. In this limit, $P^a$ are massive and they can be integrated out.
The low energy theory is the usual IYIT model with the dynamical scale $\L^4_L=m\L^3$. Confinement occurs and the effective superpotential is given by
\beq
W_{\rm eff} =\frac{1}{2} \l S_{ij}N^{ij}+X\{ {\rm Pf}(N)-\L^4_L\},
\eeq
where $N^{ij}=Q^iQ^j$ are low energy mesons, and $X$ is a lagrange multiplier.
The equation of motion of $X$ gives $\vev{N} \sim \L^2_L$, then the $F$-term of $S$ is $F^\dagger_S \sim \vev{N} \sim \L^2_L$.
Thus the SUSY is broken by the $F$-term of $S$. 

Next consider the limit $m \ll \L$ . In this case, the low energy theory can be described by mesons $N^{ij}=Q^iQ^j$, $K=P^1P^2$, and 
$L^{ia}=Q^iP^a$. The superpotential is,
\beq
W_{\rm eff}=\frac{1}{2} \l S_{ij}N^{ij} + mK -\frac{1}{\L^3}\left( K{\rm Pf}(N) + \cdots \right) ,
\eeq
where dots denote terms containing $L^{ia}$, which are unimportant for the discussion below.
The leading terms in the K\"ahler potential of the low energy theory may be of the form
\beq
K_{\rm eff} \simeq \sum_{i<j} \left\{|S_{ij}|^2 + \frac{1}{c^2|\L|^2}|N^{ij}|^2  \right\} + \frac{1}{c^2|\L|^2}|K|^2 + \cdots,
\eeq
where $c$ is a positive numerical constant. Then the potential is 
\beq
V \simeq \sum_{i<j} \left\{ |\l N^{ij}|^2 + c^2 |\L|^2 \left| \l S_{ij} - \frac{1}{2\L^3} K \e_{ijkl}N^{kl} \right|^2 \right\} 
+ c^2 | \L|^2 \left| m -  \frac{1}{\L^3} {\rm Pf}(N) \right|^2,
\eeq
where $\e_{ijkl}$ is the totally anti-symmetric tensor with $\e_{1234}=1$.
Using this potential, we can see that the minima of the potential are at $S=N=0$ with $-F^\dagger_K=c^2 |\L|^2m$ and $F^\dagger_S=0$, 
if the condition $|m/\L| \ll | \l |^2/c^2$ is satisfied. Around $|m/\L| \sim | \l |^2/c^2$, one can see that a phase transition occurs~\footnote{
In a parameter region $|m/\L| \gsim 1$, the meson description $K$ becomes worse and the quark description $P^a$ becomes better.}. 

Thus we can conclude that in the limit $m \ll \L$, the $F$-term of $S$ is zero,
while in the other limit $m \gg \L$, the $F$-term of $S$ is non-zero. Such a transition may also occur in our model.


\begin{thebibliography}{99}

\bibitem{Intriligator:2006dd}
  K.~A.~Intriligator, N.~Seiberg and D.~Shih,
  JHEP {\bf 0604}, 021 (2006)
  [arXiv:hep-th/0602239].
  
\bibitem{Kitano:2010fa}
  R.~Kitano, H.~Ooguri and Y.~Ookouchi,
  arXiv:1001.4535 [hep-th], and references therein.


\bibitem{Izawa:1997hu}
  K.~I.~Izawa and T.~Yanagida,
  Prog.\ Theor.\ Phys.\  {\bf 114}, 433 (2005)
  [arXiv:hep-ph/0501254].
  See also K.~I.~Izawa,
  Prog.\ Theor.\ Phys.\  {\bf 98}, 443 (1997)
  [arXiv:hep-ph/9704382].
  
\bibitem{Seiberg:2008qj}
  N.~Seiberg, T.~Volansky and B.~Wecht,
  JHEP {\bf 0811}, 004 (2008)
  [arXiv:0809.4437 [hep-ph]].
  
\bibitem{Jones:2008ib}
  J.~L.~Jones,
  Phys.\ Rev.\  D {\bf 79}, 075009 (2009)
  [arXiv:0812.2106 [hep-ph]].

\bibitem{Sato:2009yt}
  R.~Sato, T.~T.~Yanagida and K.~Yonekura,
  arXiv:0910.3790 [hep-ph].
  
\bibitem{Kitano:2006xg}
  R.~Kitano, H.~Ooguri and Y.~Ookouchi,
  Phys.\ Rev.\  D {\bf 75}, 045022 (2007)
  [arXiv:hep-ph/0612139].


\bibitem{Izawa:2009nz}
  K.~I.~Izawa, F.~Takahashi, T.~T.~Yanagida and K.~Yonekura,
  Phys.\ Rev.\  D {\bf 80}, 085017 (2009)
  [arXiv:0905.1764 [hep-th]].
 See also K.~I.~Izawa, F.~Takahashi, T.~T.~Yanagida and K.~Yonekura,
  Phys.\ Lett.\  B {\bf 677}, 195 (2009)
  [arXiv:0902.3854 [hep-th]].

\bibitem{Ibe:2007wp}
  M.~Ibe, Y.~Nakayama and T.~T.~Yanagida,
  Phys.\ Lett.\  B {\bf 649}, 292 (2007)
  [arXiv:hep-ph/0703110];
  M.~Ibe, Y.~Nakayama and T.~T.~Yanagida,
  Phys.\ Lett.\  B {\bf 671}, 378 (2009)
  [arXiv:0804.0636 [hep-ph]].

\bibitem{Seiberg:1994pq}
  N.~Seiberg,
  Nucl.\ Phys.\  B {\bf 435}, 129 (1995)
  [arXiv:hep-th/9411149].
  
\bibitem{Intriligator:1995au}
  K.~A.~Intriligator and N.~Seiberg,
  Nucl.\ Phys.\ Proc.\ Suppl.\  {\bf 45BC}, 1 (1996)
  [arXiv:hep-th/9509066].

\bibitem{ArkaniHamed:1997ut}
  N.~Arkani-Hamed and H.~Murayama,
  Phys.\ Rev.\  D {\bf 57}, 6638 (1998)
  [arXiv:hep-th/9705189].

\bibitem{Witten:1982df}
  E.~Witten,
  Nucl.\ Phys.\  B {\bf 202}, 253 (1982).
  
\bibitem{Affleck:1983mk}
  I.~Affleck, M.~Dine and N.~Seiberg,
  Nucl.\ Phys.\  B {\bf 241}, 493 (1984).

  
\bibitem{Izawa:2008ef}
  M.~Ibe, K.~I.~Izawa and Y.~Nakai,
  arXiv:0812.4089 [hep-ph];
  M.~Ibe, K.~I.~Izawa and Y.~Nakai,
  Phys.\ Rev.\  D {\bf 80}, 035002 (2009)
  [arXiv:0907.2970 [hep-ph]].

\bibitem{Viel:2005qj}
  M.~Viel, J.~Lesgourgues, M.~G.~Haehnelt, S.~Matarrese and A.~Riotto,
  Phys.\ Rev.\  D {\bf 71}, 063534 (2005)
  [arXiv:astro-ph/0501562].


\bibitem{Novikov:1983uc}
  V.~A.~Novikov, M.~A.~Shifman, A.~I.~Vainshtein and V.~I.~Zakharov,
  Nucl.\ Phys.\  B {\bf 229}, 381 (1983);
  M.~A.~Shifman and A.~I.~Vainshtein,
  Nucl.\ Phys.\  B {\bf 277}, 456 (1986)
  [Sov.\ Phys.\ JETP {\bf 64}, 428 (1986\ ZETFA,91,723-744.1986)];
  N.~Arkani-Hamed and H.~Murayama,
  JHEP {\bf 0006}, 030 (2000)
  [arXiv:hep-th/9707133].
  
\bibitem{Intriligator:2003jj}
  K.~A.~Intriligator and B.~Wecht,
  Nucl.\ Phys.\  B {\bf 667}, 183 (2003)
  [arXiv:hep-th/0304128].

  
\bibitem{Argurio:2009ge}
  R.~Argurio, M.~Bertolini, G.~Ferretti and A.~Mariotti,
  arXiv:0912.0743 [hep-ph].



\bibitem{Izawa:1996pk}
  K.~I.~Izawa and T.~Yanagida,
  Prog.\ Theor.\ Phys.\  {\bf 95}, 829 (1996)
  [arXiv:hep-th/9602180].
  
\bibitem{Intriligator:1996pu}
  K.~A.~Intriligator and S.~D.~Thomas,
  Nucl.\ Phys.\  B {\bf 473}, 121 (1996)
  [arXiv:hep-th/9603158].
  
  \bibitem{Nakamura}
  E.~Nakamura, Private communication.




\end{thebibliography}
\end{document}